Whitepaper for ASTRO2010

# Estimation of the Hubble Constant and Constraint on Descriptions of Dark Energy

Submitted by Lincoln Greenhill, Elizabeth Humphreys (CfA), Wayne Hu (U. Chicago), Lucas Macri (Texas A&M), David Murphy (JPL), Karen Masters (Portsmouth), Yoshiaki Hagiwara, Hideyuki Kobayashi (NAOJ), and Yasuhiro Murata (JAXA/ISAS Deputy Project Scientist ASTRO-G / VSOP-2 mission)

**Contact person:**
Lincoln Greenhill, Center for Astrophysics, 60 Garden St, Cambridge, MA 02138
greenhill@cfa.harvard.edu, phone: 617-495-7194, fax: 617-495-7345

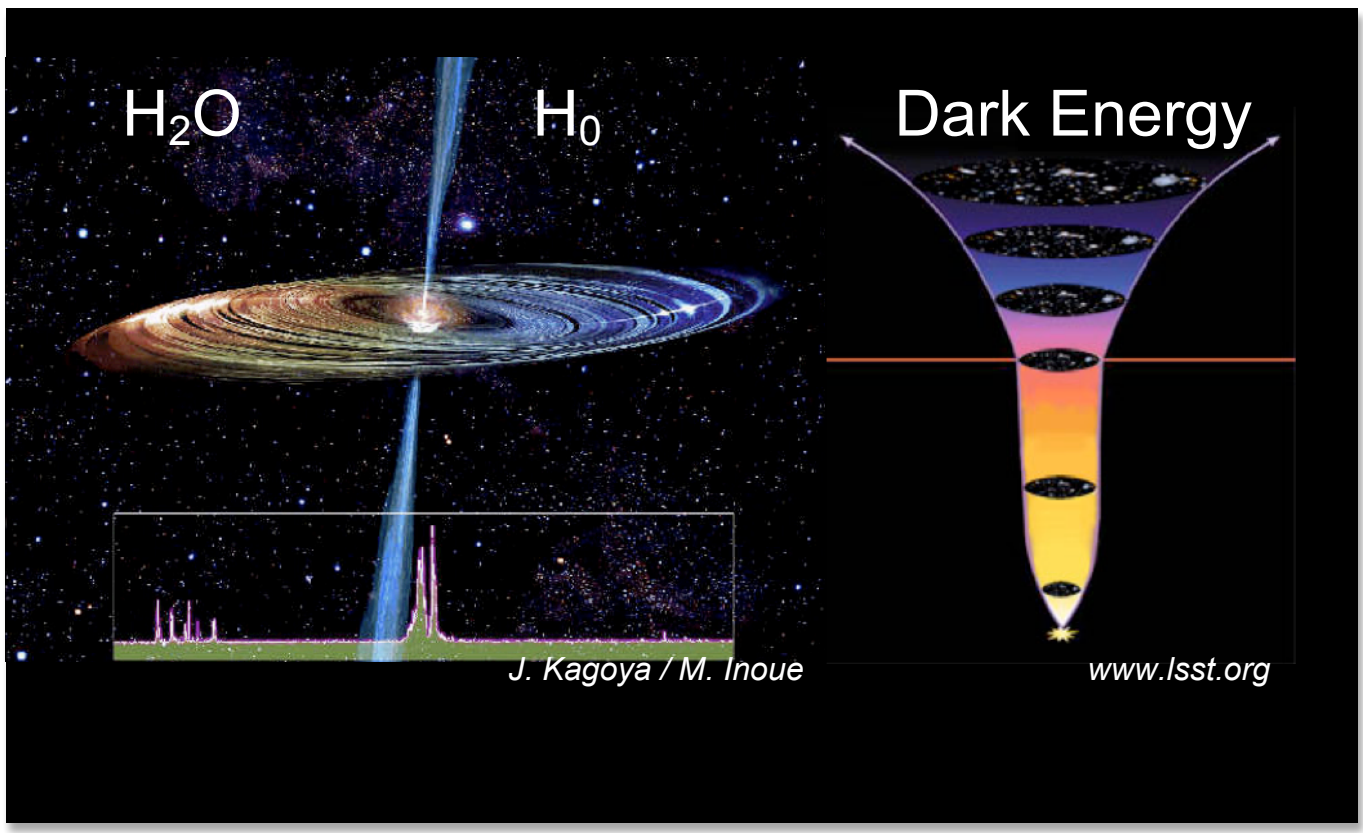

*J. Kagoya / M. Inoue*     *www.lsst.org*

**Abstract**

Joint analysis of Cosmic Microwave Background, Baryon Acoustic Oscillation, and supernova data has enabled precision estimation of cosmological parameters. New programs will push to 1% uncertainty in the dark energy equation of state and tightened constraint on curvature, requiring close attention to systematics. Direct 1% measurement of the Hubble constant (H0) would provide a new constraint. It can be obtained without overlapping systematics directly from recessional velocities and geometric distance estimates for galaxies via the mapping of water maser emission that traces the accretion disks of nuclear black holes. We identify redshifts 0.02<z<0.06 as best for small samples, e.g., 10 widely distributed galaxies, each with 3% distance uncertainty. Knowledge of peculiar radial motion is also required. Mapping requires very long baseline interferometry (VLBI) with the finest angular resolution, sensitivity to individual lines of a few mJy-km/s, and baselines that can detect a complex of ~10 mJy lines (peak) in < 1 min. For 2010-2020, large ground apertures (50-100-m diameter) augmenting the VLBA are critical, such as EVLA, GBT, Effelsberg, and the Large Millimeter Telescope, for which we propose a 22 GHz receiver and VLBI instrumentation. A space-VLBI aperture may be required, thus motivating US participation in the Japanese-led VSOP-2 mission (launch c.2013). This will provide 3-4x longer baselines and ~5x improvement in distance uncertainty. There are now 5 good targets at z>0.02, out of ~100 known masers. A single-dish discovery survey of >10,000 nuclei (>2500 hours on the GBT) would build a sample of tens of potential distance anchors. Beyond 2020, a high-frequency Square Kilometer Array (SKA) could provide larger maser samples, enabling estimation of H0 from individually less accurate distances, and possibly without the need for peculiar motion corrections.

**Thematic Science Area:** *Cosmology and Fundamental Physics*

**Goal:** a 1% estimate of $H_0$, enabling a new, robust and independent constraint on descriptions of dark energy, complementing other precision cosmology programs, e.g., Sloan Digital Sky Survey, Planck.

**Technique:** survey galaxies ($10^3$-$10^4$ 0.02<$z$<0.06) to discover $H_2O$ masers in disks around black holes in galactic nuclei; ground and space very long baseline interferometry (VLBI) and time-series spectroscopy of 10-100 masers to estimate geometric distances; combine these with recessional velocities, minus peculiar motions, to obtain $H_0$.

**Instrumentation:** Intercontinental 22 GHz VLBI arrays (VLBA) with large high-sensitivity apertures (GBT, EVLA, Effelsberg, LMT, DSN); addition of a spaceborne VLBI antenna and 30,000 km baselines; high-frequency Square Kilometer Array (HF-SKA) pathfinder.

**Opportunity in the next decade:** collaborate with the Japanese space agency (JAXA) on the VSOP-2 mission that will orbit a VLBI antenna (ASTRO-G); participation in international SKA development efforts, focusing on a high-frequency path.

## 1. Dark Energy and Curvature

The "standard" model describes a universe that is very nearly flat in geometric terms. It comprises familiar components, baryonic matter and radiation, and unseen components that are detected only indirectly, dark matter and energy. The latter two drive the evolution of the universe, which models suggest today is in transition from an epoch dominated by gravitational deceleration to an epoch dominated by acceleration due to dark energy (Riess et al. 2004) **– but what is dark energy?**

Baryonic matter comprises just 4% of the universe, in contrast to 23% for dark matter, and 73% for dark energy (Komatsu et al. 2008). Determination of cosmological model parameters with high internal consistency has been achieved through analyses of the angular power spectrum for the CMB (Dunkley et al. 2008), supplemented principally by relative distances and redshifts for type Ia supernovae (SNe) host galaxies (e.g., Kowalski et al. 2008), and the signature of baryon acoustic oscillation (BAO) for $z$ =0.2 and 0.35 among galaxies in the 2dF Redshift Survey and Sloan Digital Sky Survey (SDSS) – (Eisenstein et al. 2005; Percival et al. 2007).

Characterization of dark energy focuses on estimation of the equation of state (EOS), which is the ratio of pressure to density (p = $w\rho$). For time-invariant $w$ = –1, the EOS is consistent with a Cosmological Constant. The physics responsible would be unclear; vacuum fluctuations are a possibility, but the magnitude is many orders away from predictions of quantum theory (review by Peebles & Ratra 2003). Any other fixed or time variable value of $w$ would require (still) more exotic physics. Quintessence is one candidate cause for dynamical dark energy for which $w$ is < –1/3 and time variable (Chongchitnan & Efstathiou 2007, references therein).

Observational constraints set by CMB, SN, and BAO data are consistent with $w$ = –1, but uncertainties are large enough that a Cosmological Constant is not a certainty. Figure 1 shows analysis for constant $w \neq$ -1. Admitting that it may be time variable, Komatsu et al. (2008) obtain –1.33 < $w$ < –0.79 at $z$=0, with 95% confidence, for a flat universe. *Although w is close to –1, and though this outcome may appear most plausible, w is not necessarily –1.*

Even a small deviation, $w \neq$ –1, would have broad physical implications, including weakening constraint on curvature, $\Omega_k$. Expectations from Inflation theory are that measurable curvature would be < $O(10^{-5})$. Any excess would be noteworthy (e.g., Knox 2006). In the context of a Cosmological constant, current CMB and principally BAO data constrain curvature well (-0.018 < $\Omega_k$ < 0.007 at 95% confidence), where BAO data contribute an important absolute measure of angular-diameter distance for $z$ = 0.35 (Eisenstein et al. 2005; Percival et al. 2007). Joint analysis of CMB, SN, and BAO data for constant $w \neq -1$ yields $-0.018 < \Omega_k < 0.008$ (Fig. 1; Komatsu et al. 2008). Admission of a time-variable EOS weakens constraint, probably by a further 50% (cf., Wang & Mukherjee 2007) **– is the universe not flat afterall?**

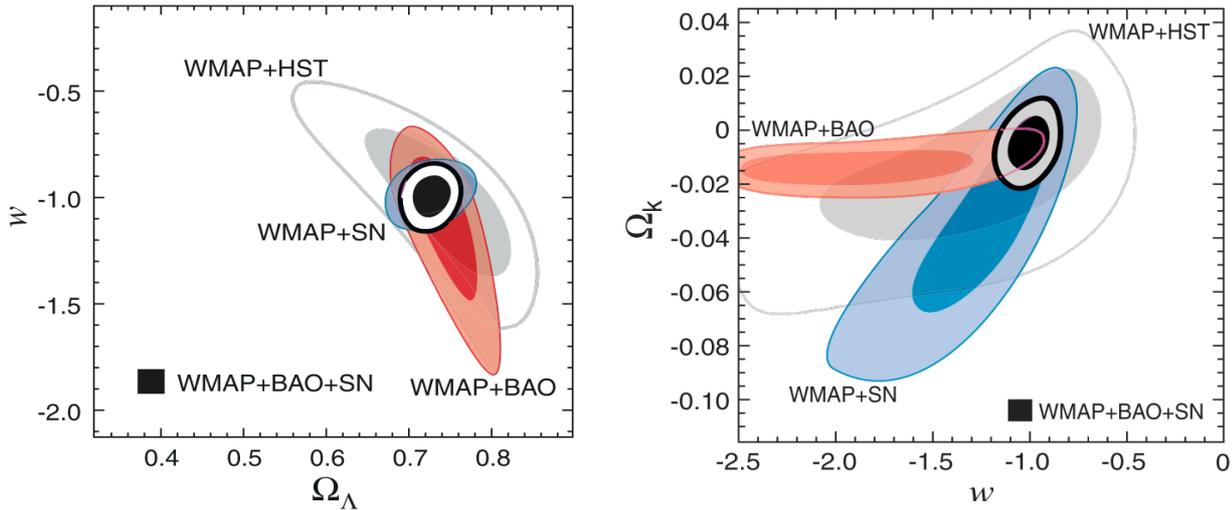

**Figure 1** – Estimation of cosmological parameters in analysis of CMB, SN, and BAO data (Komatsu et al. 2008). Contours are 1 and 2$\sigma$; $w \neq 1$ but it is constant. Parameters not shown have been marginalized. (**left**) Flat universe. (**right**) Non-flat universe (open: $\Omega_k > 0$; closed: $\Omega_k < 0$). Constraints on parameter relax considerable if flatness is not required. Admitting time variability of $w$ exacerbates this further (cf., Wang & Mukherjee 2007). The Freedman et al. (2001) estimate of $H_0$ is applied here. We focus on the contribution of a future estimate with ~10x smaller uncertainty (1%).

Techniques that provide high-accuracy measurements of absolute distance are sought-after but rare. Direct measurement of $H_0$ and BAO are two of a kind. Pursuing the former establishes the distance scale at $z = 0$, narrows uncertainties for $w$ and $\Omega_k$ (Hu 2005; Ichikawa & Takahashi 2008), delivers a constraint that is fundamental, and for the techniques discussed in Section 3, provides a measurement that is genuinely independent in terms of calibration and systematics.

## 2. Constraints on Dark Energy and Curvature from $H_0$

Joint analyses to constrain the EOS, curvature, and other parameters (e.g., neutrino mass; Ichikawa 2008) include priors on $H_0$, derived from systematic and methodical studies of Cepheid variable stars in nearby galaxies (e.g., 72 ±3(random) ±7(systematic) km s$^{-1}$ Mpc$^{-1}$; Freedman et al. 2001). However, the accuracy and consistency of other data types (CMB, BAO, SNe) has progressed sufficiently that a Cepheid–based estimate of $H_0$ with 10% uncertainty adds little to global solutions of cosmological parameters (Fig. 1). Recent study by Macri et al. and Riess et al. (in prep) has established a 5% uncertainty using a sample of well-calibrated SNe in nearby galaxies, Cepheids within a subset of these, and Cepheids in the galaxy NGC4258, to which a geometric distance is available (see Section 5). These studies have led to better constraints on $w$, but further progress via this traditional route will be unlikely until well into the JWST era and will be limited by remaining systematic uncertainties in the Cepheid and SN distance scales, and uncertainty in the distance to NGC 4258. Higher *precision* estimates of $H_0$ have been inferred from CMB data under the assumptions $w = -1$ and $\Omega_k = 0$ (e.g., Spergel et al. 2007), and these agree with the "Cepheid estimates," though the assumptions may be problematic.

Hu (2005) demonstrates the coupling of EOS parameters ($w$ at $z=0$ and the derivative $w'$) to $H_0$ and concludes that percent accuracy in the latter could enable sensitive discrimination among models (Fig. 2). Ichikawa & Takahashi (2008) undertake formal analysis of the marginalized probability distributions for current CMB, BAO, and SN datasets (Fig. 3), adding 3% priors on $H_0$, with and without curvature and dynamical dark energy. Their choice of priors on $H_0$ (62 and 72 km s$^{-1}$ Mpc$^{-1}$) covers the range of recently published values (e.g., Macri et al. 2006 and di Benedetto 2008). The cumulative results demonstrate the reduced leverage of CMB, BAO, and SN data alone for non-flat and dynamical models, the importance of the actual $H_0$ priors, and an indicative scaling where 1% uncertainty in $H_0$ broadens the range of models for which combined analyses may result in $O(1\%)$ constraint on EOS parameters (though a formal probability analysis remains to be done.)

Direct estimation of the expansion rate today, i.e., $H_0$, with 1% accuracy from distance measurements for a sample of anchor galaxies naturally complements what is anticipated for new BAO studies of expansion rate at many redshifts. Current BAO accuracy is ~ 4% at $z = 0.35$ (Eisenstein et al. 2005; Percival et al. 2007), and the SDSS Baryon Oscillation Spectroscopic Survey (BOSS) will enable 1.1% accuracy at $z \sim 0.6$ and 1.5% at $z = 2.5$. Considering identified biases in technique and models, overall accuracy of 1-2% is anticipated for the expansion rate, $H(z)$ (Eisenstein et al. 2007a, b; Seo et al. 2008; and e.g., Doran et al. 2007; Linders & Robbers 2008). A 2009-2014 timeline for the BOSS effort sets the time scale on which high-accuracy estimation of $H_0$ is required. Analyses of data from the Planck mission, which will deliver sub-percent uncertainties in description of the CMB (e.g., summary by White 2006), are due in the same period.

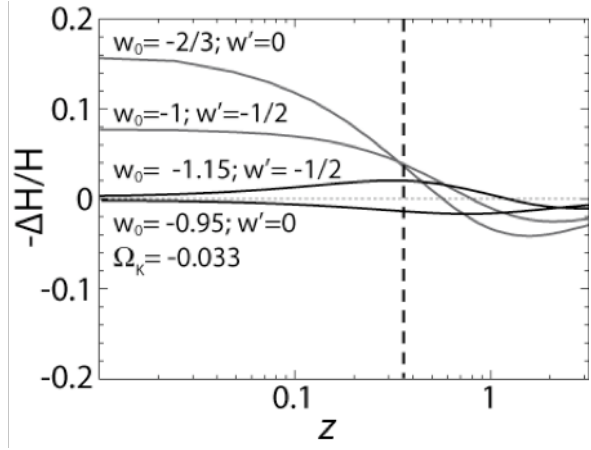

**Figure 2** – Variation of the Hubble parameter with $z$ for four models including a fixed or variable EOS, and flat or non-flat geometry. Assumed critical densities are as in Spergel et al. (2003). The results demonstrate sensitivity of $H_0$ to the form of the EOS. Conversely, tight constraint on $H_0$ would affect estimates of $w$ at $z = 0$ and $w'$. The dashed line indicates the redshift of spectroscopic BAO measurement in Eisenstein et al. (2005), where variation in $H$ is much smaller than variation in $H_0$ for this specific set of models. Adapted from Hu 2005.

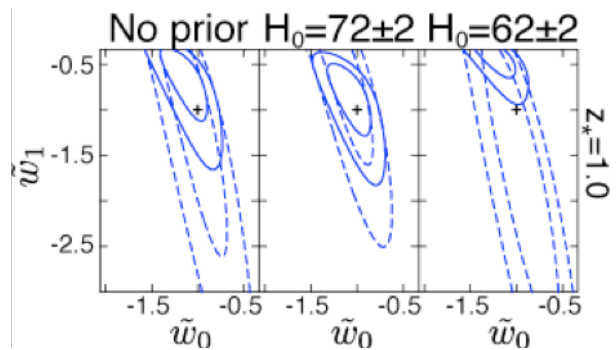

**Figure 3** – Dark energy constraints for flat (solid) and non-flat (dashed) cosmologies, demonstrating sensitivity to priors on $H_0$. In this model, $w$ varies linearly from a fixed $\varpi_1$ at $z > 1$ to $\varpi_0$ at $z=0$. Contours denote 1 and 2$\sigma$. Crosses indicate $\varpi_0, \varpi_1 = -1$, corresponding to $w \equiv -1$. CMB data are from Spergel et al. (2007), SN data from Davis et al. (2007), and BAO data from Eisenstein et al. (2005). Adapted from Ichikawa & Takahashi (2008).

## 3. Estimating $H_0$ Directly

Few direct techniques can deliver high-accuracy estimates of $H_0$. High-angular resolution study of $H_2O$ maser sources that lie in the accretion disks of AGN ("disk masers") can contribute a direct, geometric measurement, built on distance estimates to individual anchor galaxies. The strength of these estimates lies in a well-defined geometry for the underlying astrophysical system, with relatively few parameters. Moreover, the systematic errors are independent of those for other techniques used in cosmological parameter estimation (e.g., standard candles).

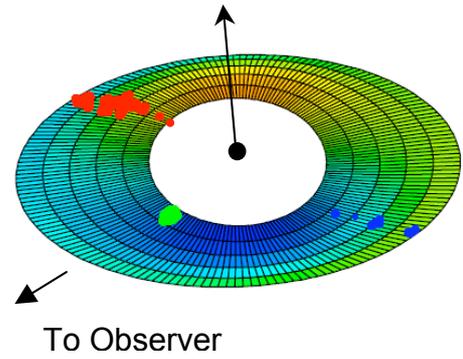

**Figure 4**–Maser data and accretion disk model for archetype NGC 4258. High-velocity water masers observed with VLBI are shown along the diameter perpendicular to the line of sight; systemic masers are on the nearside. The mesh represents the best-fitting warp model (Humphreys et al. 2008, and references therein.

The accretion disks of massive black holes in galactic nuclei, when viewed nearly edge-on, are outlined by high-intensity (maser) emission that arises from $H_2O$ molecules in disk gas warmed by X-ray irradiation (e.g., Neufeld et al. 1994). The resulting $H_2O$ spectrum comprises Doppler components close to the systemic velocity of the central engine and features symmetrically offset by the rotation speed(s) of the disk (a.k.a. systemic and high-velocity maser emission). When angular and velocity structure can be resolved using Very Long Baseline Interferometry (VLBI), with sub-milliarcsecond angular resolution, it is possible to infer disk geometry and central masses ($10^6$-$10^8$ M☉; Tilak et al. 2008, and references therein) with little more than Newton's Law. Maser-VLBI is the only known means by which to directly image the structure of accretion disks at radii « 1 pc; no other imaging technique reaches as deeply into nuclei. Distances are obtained from 3-D dynamical models of disks (Fig. 4) that are fit in $\chi^2$ minimizations to VLBI data and centripetal accelerations of maser material, which are evident from the secular change of line-of-sight velocities observed in single-dish time-series spectra of systemic emission (e.g., Herrnstein et al. 2005) and references therein).

In order to understand the dependence of distance (D) on different observables, we formulate an approximate distance calculation in terms of systemic accelerations and what can be read from an observed maser position-velocity diagram. We treat systemic and high-velocity masers separately because they may occur at different radii. The Keplerian high-velocity rotation curve $v_{rot}^2 \propto (M/D\theta_{high})$ may be read to $0^{th}$ order from a position-velocity diagram, where $v_{rot}$ is the LOS velocity of high-velocity emission at angular radius $\theta_{high}$. This yields the distance-dependent enclosed mass ($M/D$). Linear position-velocity (P-V) gradients in systemic emission are commonplace, corresponding to $\zeta = dv_{los}/db_{low} \propto (M/D\theta^3_{low})^{1/2}$. This provides $\theta_{low}$, the mean disk radius from which systemic emission originates, in angular units. (The impact parameter from the disk center is $b$.) Since $a_{los} \propto (M/D)(1/D\theta^2_{low})$, we can write $D \propto a_{los}^{-1} [\theta_{high}v_{rot}^2]^{1/3} \zeta^{4/3}$. Clearly, accurate measurement of $a_{los}$ and $\zeta$ is a priority.

## 4. Current Sample Size and Detection Rates

Disk masers are rare; after surveys of $O(3000)$ nuclei, there are ~ 100 known masers in AGN and ~ 40 originating in disks. About 20 of these show "triple" spectra (i.e., displaying systemic and high-velocity emission in three line complexes), and

~10 appear to be candidates for distance measurement, depending on the collecting area available for VLBI. Five of these lie at $z > 0.02$; we discuss the origin of this lower limit in Section 5. Among type-2 Seyferts and LINERs, detection rates among surveys for disk masers are 3-5% for $0.02 < z < 0.05$ and at a sensitivity ($3\sigma$) of ~ 10 mJy in a 1 km s$^{-1}$ spectral channel (e.g., Braatz et al. 2004; Kondratko et al. 2006). However, the true incidence rate of disk maser emission is not known because the census of AGN is incomplete. Among optically normal (S0, Sa, Sb) nuclei, the disk-maser detection rate for a largely complete survey ($z < 0.016$; $M_B > -19.5$) is ~ 10% as large for a sensitivity ($3\sigma$) of ~ 15 mJy (Braatz & Gugliucci 2008). Presumably, these are unrecognized AGN.

The disk masers most likely to yield high-accuracy geometric distances meet six criteria. Each should: (i) display a triple spectrum of narrow lines; (ii) have broad line complexes (»100 km s$^{-1}$) that provide good leverage on estimation of rotation curves ($M/D$) and systemic P-V gradients ($\zeta$); (iii) exhibit measurable centripetal accelerations ($a_{los}$); (iv) have Doppler components strong enough to enable VLBI imaging throughout the velocity ranges of each complex; (v) exhibit Keplerian rotation in thin disk structures; and (vi) lie at great enough distances that residual peculiar radial motion is small relative to the recessional velocities. The need to resolve disk structure well with VLBI favors study of nearby sources, while reduction in fractional peculiar motion favors distant sources.

## 5. Is 1% Uncertainty in $H_0$ Feasible?

In summary, 1% accuracy in $H_0$ may be sought via two routes: measurement of high-accuracy distances to a small number of disk masers, e.g., 10 estimates individually good to 3%, and measurement of low accuracy distances to a large sample of disk masers, e.g., > 100 estimates individually good to 10% or less. We focus on $0.02 < z < 0.06$ and the first route, because in the next half decade, the number of disk masers with high-quality distances is likely to remain below 100 (see Section 6).

Naively, $H_0$ can be obtained from a single galaxy recessional velocity divided by a corresponding maser distance. The distance uncertainty for the archetypal NGC4258 disk, which is bright, large in angle, clean (e.g., thin, Keplerian), and richly studied is ~ 3% (Herrnstein et al. 1999, 2005; Humphreys et al., 2008). We adopt this as a conservative floor for each more distant system, where linear resolution is lower, and the deprojection of maser positions onto the face of the disk is less accurate. This particularly affects systemic masers, which lie at different radii but along very similar lines of sight; error is equivalent to inaccurate measurement of $\zeta$ ($D \propto \zeta^{4/3}$). The resulting uncertainty in radius also limits interpretation of $a_{los}$ for individual maser components, and directly affects uncertainty in distance ($D \propto a_{los}$).

For a sample of N anchors, uncertainty in $H_0$ may scale as ~ $N^{-0.5}$ (for uniform uncertainties) because errors for individual distance measurements are uncorrelated; they depend principally on the distribution of Doppler components across the disk face and details of disk geometries, which differ from galaxy to galaxy. However, the peculiar radial motions of galaxies *are* partially correlated and increase scatter in the Hubble relation. The sample size needed to achieve a given total uncertainty grows as the quadrature sum of fractional error in distance and fractional peculiar motion. Hence, when $\Delta D/D$ is small, correction for peculiar motion is a priority. Most are < 500 km s$^{-1}$ outside of clusters (Springob et al. 2007), and galaxies closer than $cz = 500(\Delta D/D)^{-1}$ are affected. For the most distant known disk maser ($z=0.059$), this *alone* corresponds to a 3% error in the Hubble parameter.

Extension of maser studies to recessional velocities at which peculiar motions can be ignored entirely is desirable but probably impractical; $z \sim 0.06$ is an approximate limit. This is where the smallest known maser-disk inner diameter (~0.2 pc) subtends less than one half of a ground-based VLBI resolution element ($\lambda$1.35 cm and 8000 km baseline). Further reduction in angular extent critically affects measurement of $\zeta$ and interpretation of $a_{los}$. Larger disks are known (e.g., Tilak et al. 2008), but observable acceleration declines sharply with radius ($r^{-2}$). Were the disk diameter 2 pc for the relatively massive central engine in NGC4258 ($4\times10^7$ M☉), the acceleration would be < 0.1 line widths per year and below useful measurement limits (Herrnstein et al. 1999).

Minimizing the impact of peculiar motions on estimation of $H_0$ requires: (i) distances to maser hosts distributed broadly over the sky, particularly those in "quiet" portions of the Hubble flow; (ii) subtraction of cosmic flow models, (iii) avoidance of regions where motions are large and poorly quantified (e.g., toward the Great Attractor), and (iv) use of barycenter velocities for maser hosts that are in rich clusters. Using Tully-Fisher distances to estimate peculiar velocities and construct a multi-attractor model inside $z \sim 0.02$, Masters (2005) obtained a 160±20 km s$^{-1}$ RMS residual. Erdogdu et al. (2006) report a somewhat less direct error estimate, a scatter of 100-150 km s$^{-1}$ around a flow model derived from galaxy recessional velocities (as opposed to peculiar velocities), up to $z \sim 0.053$. Adopting a characteristic 160 km s$^{-1}$ uncertainty for corrected recessional velocities, we suggest the maximum contribution to the error budget for an anchor will be 2.6%, at $z$=0.02, and below the 3% limit. As indicated in item (iv), substantial peculiar motion due to internal cluster dynamics will be a concern for some anchors. For clusters with > $O(10)$ members, barycenter velocities may be used instead of individual galaxy recessional velocities; we anticipate related uncertainties « 200 km s$^{-1}$ (e.g., Crook et al. 2007).

## 6. What is Required?

*Mapping*: High-quality ground-VLBI imaging of each maser disk requires detection of one or a few emission features in an atmospheric coherence time (~ 45$^s$). This imposes a severe sensitivity limit, and most known, distant disk masers are too weak to study with the VLBA. A backbone of **50-100m class apertures** augmenting the **VLBA** can boost sensitivity by > 3x, enabling follow-up of most sources discovered in the single-dish surveys described earlier. Maser-VLBI studies in the last few years have chiefly relied upon a minimal configuration: GBT, VLBA. In contrast, at least three, and preferably four, large apertures are required. The GBT, EVLA, MPIfR 100m, and NASA/DSN 70m antennas (when not tracking missions) will be ongoing critical resources (2010-2020). Outfitting the US/Mexican **Large Millimeter Telescope** (LMT) for $\lambda$1.3cm VLBI operation, and nascent 8x **expansion in the bandwidth of VLBI hardware and facilities** are top short-term development priorities. Addition of a **space-VLBI element** (and development of necessary calibration techniques) is the top mid-term priority. **US participation in the Japanese-led VSOP-2 mission** (launch 2013) would enable 3-4x higher angular resolution. This would reduce systematic uncertainty in deprojection of systemic maser emission and improve control over systematics that can affect distances (e.g., spiral disk structure, non-point symmetric warps, radius-dependent eccentricity). A 5x reduction in distance uncertainty may be achievable. A **high-frequency SKA pathfinder** (HF-SKA) that could augment the large-aperture backbone would be the top long-term development priority (2010-2020). Eventual SKA performance is described in Morganti et al. (2004), enabling scaling.

*Surveys:* Large-N single-dish surveys of nuclei are needed to increase the number of known maser disks at *z* > 0.02. The launch of VSOP-2 and completion of BOSS and Planck data analyses suggest a target of 2013. The number of disk masers required depends on the accuracy achieved in distance measurement efforts: 3% requires *O*(10) targets at *z* > 0.02; 10% requires *O*(100) targets. For a survey of type-2 AGN, $10^3$-$10^4$ galaxies may need to be observed. Among normal nuclei $10^4$-$10^5$ may be needed. **The GBT will be a necessary ongoing resource**. The next most sensitive monolithic antennas (MPIfR, NASA/DSN) are 6x less efficient in observing time for comparable instrumentation. (Development work for one DSN antenna is on going at JPL and CfA, but antenna time is strongly limited.) For $10^4$ targets, a GBT-only survey would require ~ 2500 hours equivalent to dedicated observing for about two fall/winter months per year for two years at the GBT, achieving ~ 10 mJy sensitivity ($3\sigma$) over 1 km s$^{-1}$. This would be 5x larger than recent/ongoing programs and considerably more concentrated. Some load could be diverted if a wide-field-of-view multiple feed systems were technically practical and able to see multiple targets at a time, e.g., in dense clusters. For extension of survey work to more targets a **HF-SKA or pathfinder** that meets or exceeds GBT performance at 22 GHz is the only identifiable long-term option.

*Peculiar Velocities:* Priority development is needed for statistical methods to deal with (i) sparse and noisy velocity-field data and (ii) comparison of flow models with peculiar velocity measurements. Large, homogenous peculiar velocity surveys (e.g., Springob et al. 2007; Masters et al. 2008) and an all-sky redshift survey (2MRS) exist, but the best means to obtain accurate results at up to *z* ~ 0.06 are not yet certain. Planned large all-sky surveys (e.g., LSST) could provide the sample for a deep redshift survey by fast telescopes equipped with multi-object spectrographs. However, additional measurement of peculiar velocities would require more accurate calibration of secondary distance indicators than has been demonstrated.